\documentclass[12pt,preprint]{aastex}

\shorttitle{Polarization of GRB 020405}
\shortauthors{Bersier et al.}

\begin{document}

\title{The Strongly Polarized Afterglow of GRB 020405\altaffilmark{1}}

\author{D.~Bersier\altaffilmark{2}, 
B.~McLeod\altaffilmark{2},
P.~M.~Garnavich\altaffilmark{3}, 
M.~J.~Holman\altaffilmark{2},
T.~Grav\altaffilmark{2}, 
J.~Quinn\altaffilmark{3},
J.~Kaluzny\altaffilmark{4}, 
P.~M.~Challis\altaffilmark{2},
R.~G.~Bower\altaffilmark{5}, 
D.~J.~Wilman\altaffilmark{5},
J.~S.~Heyl\altaffilmark{2,6}, 
S.~T.~Holland\altaffilmark{3},
V.~Hradecky\altaffilmark{2}, 
S.~Jha\altaffilmark{2},
K.~Z.~Stanek\altaffilmark{2}}

\altaffiltext{1}{Based on data from the MMT Observatory, a joint
facility of the Smithsonian Institution and the University of Arizona,
the 1.8m Vatican Advanced Technology Telescope, the Magellan 6.5m
Walter Baade telescope and the 2.5m DuPont telescope.}
\altaffiltext{2}{Harvard-Smithsonian Center for Astrophysics, 60
Garden St., Cambridge, MA 02138}
\altaffiltext{3}{Department of Physics, University of Notre Dame,
Notre Dame, IN 46556-5670}
\altaffiltext{4}{Copernicus Astronomical Center, Bartycka 18,
PL-00-716, Warsaw, Poland}
\altaffiltext{5}{Department of Physics, University of Durham, South
Road, Durham DH1 3LE, UK}
\altaffiltext{6}{Chandra Scholar}

\email{dbersier@cfa.harvard.edu, 
bmcleod@cfa.harvard.edu,
pgarnavi@miranda.phys.nd.edu,
mholman@cfa.harvard.edu,
tgrav@cfa.harvard.edu,
jquinn@miranda.phys.nd.edu,
jka@camk.edu.pl,
pchallis@cfa.harvard.edu,
R.G.Bower@durham.ac.uk,
d.j.wilman@durham.ac.uk,
jheyl@cfa.harvard.edu,
sholland@nd.edu,
vhradecky@cfa.harvard.edu,
saurabh@cfa.harvard.edu,
kstanek@cfa.harvard.edu}

\begin{abstract}

We report polarization measurements and photometry for the optical
afterglow of the gamma-ray burst GRB 020405.  We measured a highly
significant 9.9\% polarization (in $V$ band) $1.3\;$days after the
burst and argue that it is intrinsic to the GRB.  The light curve
decay is well fitted by a $t^{-1.72}$ power-law; we do not see any
evidence for a break between 1.24 and $4.3\;$days after the burst.  We
discuss these measurements in the light of several models of GRB
afterglows.

\end{abstract}


\section{Introduction}
\label{sect_intro}

It is now commonly accepted that the radiation from gamma-ray bursts
(GRB) afterglows comes from synchrotron radiation (e.g. M\'esz\'aros
\& Rees 1997). Since synchrotron emission is strongly polarized
(e.g. Rybicki \& Lightman 1979), one expects to see some level of
polarization in GRB afterglows.  Several models make predictions
regarding the amount of polarization one might see in a GRB afterglow
(e.g. Gruzinov \& Waxman 1999; Gruzinov 1999; Ghisellini \& Lazzati
1999; Sari 1999). If the magnetic field is globally random but with a
large number of patches ($\approx 100$) with a coherent magnetic
field, the polarization may be high, particularly in early times,
depending on the number of patches (Gruzinov \& Waxman 1999; Gruzinov
1999). Another alternative is that the field is globally symmetric
(with no net polarization) but viewing the jet off-axis will break the
symmetry and a significant polarization may result (Ghisellini \&
Lazzati 1999; Sari 1999). Microlensing could also explain a
temporarily high polarization (Loeb \& Perna 1998; Garnavich, Loeb \&
Stanek 2000).

On the observational side, polarization has been measured in several
afterglows. \citet{cov99} first detected 1.7\% polarization for GRB
990510 about $0.77\;$days after the burst, and \citet{wij99} reported
about the same amount of polarization for this burst $0.86\;$days
after the burst; the polarization angle being the same within the
error bars. Several measurements for GRB 990712 indicated a
polarization variability \citep{rol00}, from $2.9\% \pm 0.4\%$ to
$1.2\% \pm 0.4\%$ within $0.25\;$days while the angle did not appear
to change.

GRB 020405 was seen by several spacecrafts on 2002 April 05.028773
(UT) and IPN triangulation gave a tentative position of the burst
\citep{hurley}.  Prompt optical observations \citep{pricea} revealed a
new optical source at the position $\alpha_{2000} = 13^h 58^m 03{\fs}
1, \delta_{2000} = -31{\degr} 22\arcmin 22\arcsec$, which turned out
to be quickly fading (Hjorth et al. 2002; Price et al 2002b; Covino et
al 2002a). A redshift of $z=0.695$ was obtained with emission lines of
the likely host galaxy \citep{masetti}, later refined to $z = 0.6898
\pm 0.0005$ \citep{pricec}.  Polarization observations were reported
by Covino et al. (2002a, 2002b, 2002c).

\section{Observations}

\subsection{Photometry}
\label{sect_phot}

Our observations started on 2002 Apr~6.268 UT ($1.25\;$days after the
burst) with the Vatican Advanced Technology Telescope (VATT), where we
obtained $BVRI$ data.  Soon afterwards we also obtained $V$-band
polarization data (see Sect.~\ref{sect_pol}) with the 6.5m MMT. We
continued to monitor the optical transient (OT) on the VATT and on the
MMT for three more nights. 
We obtained later-time data on the Magellan 6.5m Baade telescope with
the Magic instrument on Apr~14 ($V$) and with LDSS2 on Apr~18
($R$). We also obtained a deep $R$-band exposure on the 100'' du Pont
telescope at Las Campanas on May~3. Several standard fields
\citep{landolt} have been obtained on May~2 to calibrate the data. Our
calibration is in good agreement with that reported by Simoncelli et
al. (2002), except for the $V$-band where our data are slightly
brighter by $\sim 0.08$ mag.  On the nights of May~15 and 16
additional unpolarized $V$-band data were taken with the 6.5m MMT to
be used as the reference image for the image subtraction software (see
Table~\ref{tbl_obs} for a summary of our data).


For all images the photometry has been extracted using DoPHOT
\citep{sms93} in fixed position mode. We made a template image with
several images from the first night where the optical afterglow is
clearly visible. However there is always the concern that PSF
photometry is affected by the underlying galaxy and we found that this
is indeed the case here (see below).  For this reason we used the
image subtraction program ISIS \citep{al00} for the $R$-band data;
this gave superior results.

\begin{figure}
\plotone{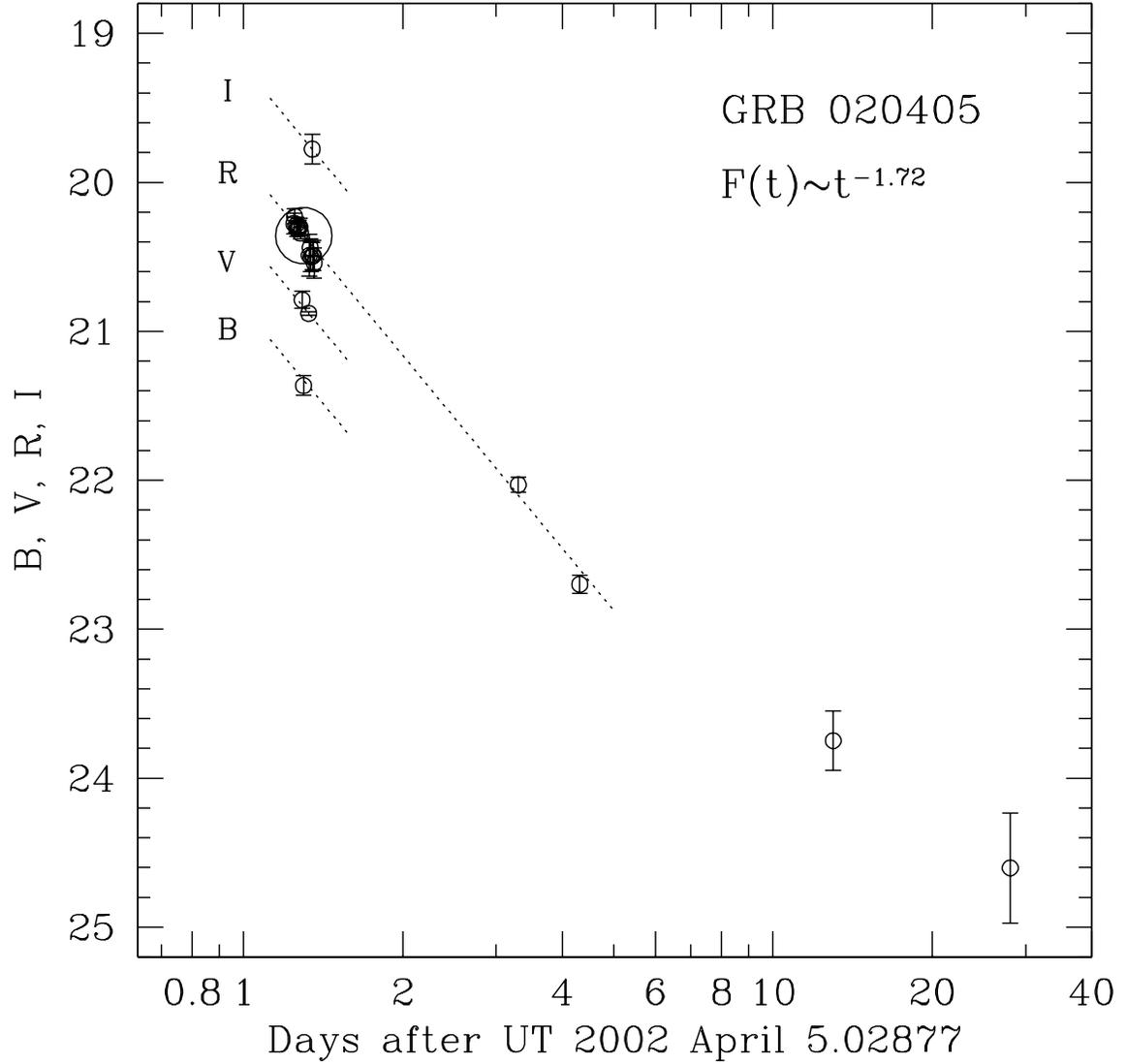}
\caption{The light curve of the optical afterglow of GRB 020405. The
power law fit is based on early (day$<5$) $R$-band data. The photometry
for these data has been obtained with image subtraction.
\label{fig_lc}}
\end{figure}

Our $BVRI$ data are plotted in Fig.~\ref{fig_lc}.  The time evolution
of the GRB is described by a power-law $F_{\nu}(t) \sim t^{-\alpha}$.
Using only ISIS results for $R$-band data obtained before Apr~10, we
obtain a slope of $\alpha=1.72$. When using all the available data
(including photometry obtained with DoPHOT) the slope was
significantly lower ($\alpha=1.3$).  This shows that the PSF
photometry in all passbands is significantly affected by the galaxy.
An important thing to note is that we do not see any evidence of a
break in the light curve between 1.24 and $4.3\;$days after the burst.
Masetti et al. (2002b) find no suggestion of a break in the light
curve up to $\sim 10\;$days after the GRB.

\begin{figure}
\plotone{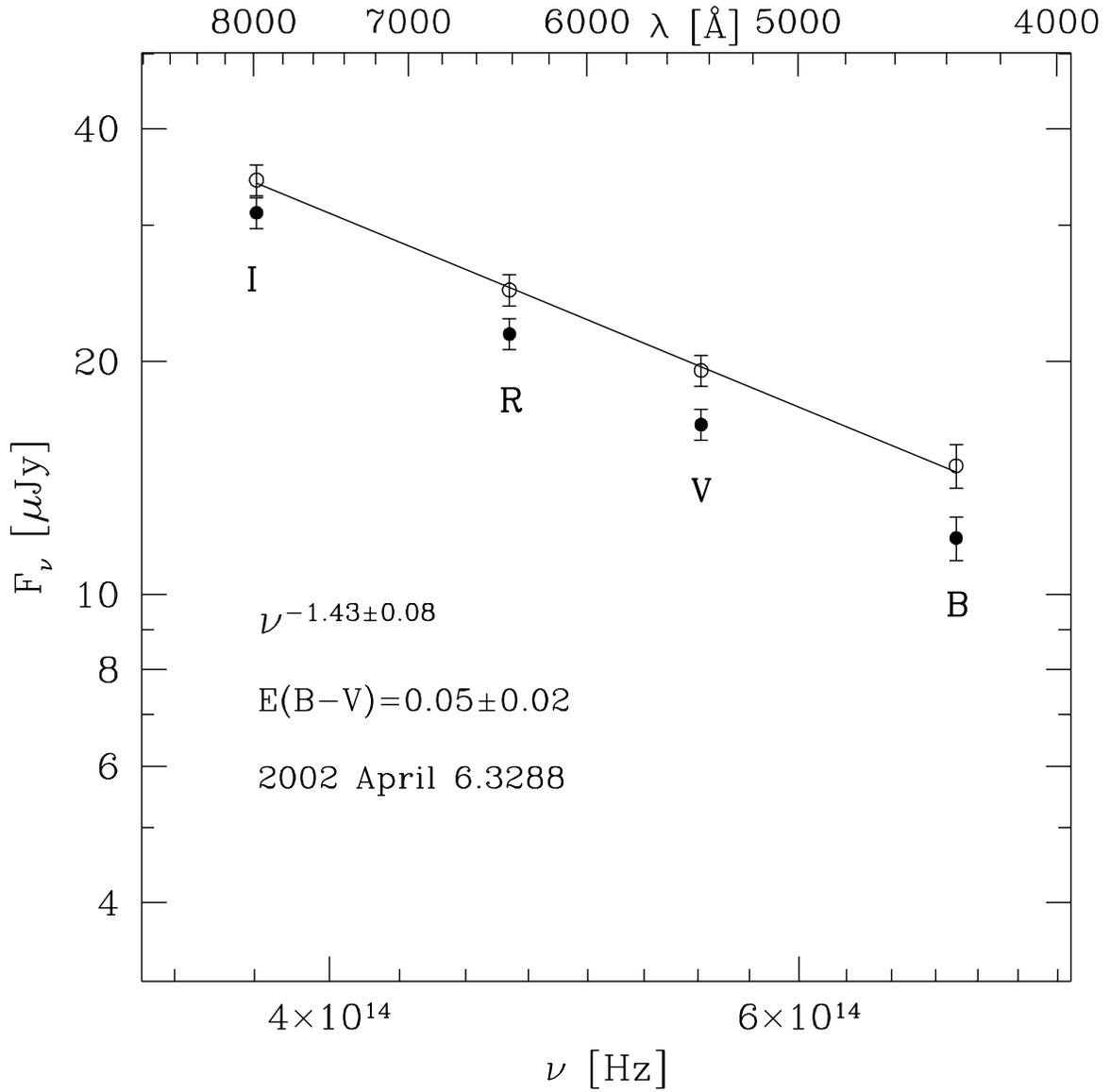}
\caption{Broad-band spectrum of the afterglow 1.3 day after the burst.
Dots indicate measurements; circles are for measurements
corrected for foreground reddening.
g\label{fig_spectrum}}
\end{figure}

Once the slope of the decay is known we can interpolate the flux in
each passband to a common time and determine the broad-band
spectrum. We chose a common date $1.3\;$days after the burst, when
most of our early data were taken, and when the galaxy's contribution
to the observed flux was smallest. The foreground reddening is $E(B-V)
= 0.054\;$mag \citep{sfd98}. We used the photometric zero points from
\citet{fsi95} to transform the magnitudes to fluxes.  The resulting
broad-band spectrum is presented in Fig.~\ref{fig_spectrum}. After
having corrected for reddening, we fitted a power law to the spectrum;
the resulting index is $\beta=1.43 \pm 0.08$.


\subsection{The Polarization Data}
\label{sect_pol}

Polarization data for the OT were obtained with the MMT 6.5m telescope
using the MiniCam imager (McLeod et al., in preparation) with a set of
four Melles-Griot polarization filters (McLeod \& Garnavich, in
preparation) on 2001 Apr~6.3 UT, starting $\approx 1.3\;$ days after
the burst. We employed $2\times2$ binning resulting in a pixel scale
of $0\farcs091$ per pixel.  Two repetitions of the following sequence
were obtained for the OT through the $V$-band filter: CLEAR, POL0,
POL45, POL90, POL135.  The polarization data were obtained in less
than ideal observing conditions, through thin clouds, high airmass of
$\sim2.2$ and rather poor seeing of $>1.6''$.  Also, on the nights of
May~15 and 16 UT additional unpolarized $V$-band data were taken with
the same instrument to be used as the reference image for the image
subtraction software.

The data were processed in two ways: with the PSF-fitting code DAOPHOT
(Stetson 1987, 1992) and with the ISIS image subtraction package
(Alard \& Lupton 1998; Alard 2000). They gave consistent polarization
measurements, but with the errors from the image subtraction being
about half of those from the PSF-fitting, which was not unexpected
(see e.g. Mochejska et al.~2002).  The ISIS reduction procedure
consists of the following steps: (1) transformation of all frames to a
common $(x,y)$ coordinate grid; (2) construction of a reference image
from several best exposures; (3) subtraction of each frame from the
reference image; (4) selection of stars to be measured; (5) extraction
of photometry from the subtracted images.  This approach allows to use
all the information in the image to extract the best photometry. It
also allows us to remove the contribution from any constant,
non-polarized flux (host galaxy) and to check for possible systematic
errors by extracting photometry for other stars in the field.  For the
reference image we have used a combination of the best images from
April and also images from May 2002. Using the images obtained after
the GRB has faded (from May) has the advantage that we not only
measure the flux difference through the different polarization
filters, but we also measure the total flux of the OT, thus allowing a
more precise measurement of the fractional polarization.

The polarization filters were calibrated under photometric conditions
on 2002 March 12 UT by taking V-band images of two polarization
standards (Schmidt, Elston, \& Lupie 1992).  The transmission
variations among the four polarizing filters were normalized using
images of the unpolarized standard G191B2B (Schmidt et al. 1992). Note
that this normalization happens automatically during ISIS reductions
described above.  For BD+59\degr389 we derive P=0.071,
$\theta=100\degr$ compared with the accepted value of P=0.067,
$\theta=98\degr$.  For HD155197 we derive P=0.046, $\theta=105\degr$,
compared with P=0.043, $\theta=103\degr$.  Hence, we see calibration
errors of $\Delta P/P=0.06$ and $\Delta\theta=2\degr$.  These errors
may be systematic and correctable. However, because the two
calibration stars have nearly the same polarization position angle we
have no information on how the error changes as a function of position
angle.  Therefore we do not attempt to correct our measured
polarization, but instead add the calibration error in quadrature to
our quoted uncertainty.

The reduction procedure described above resulted in the detection of a
strong, $\sim 10\%$ polarization 1.3 days after the burst, with each
individual measurement having $1\sigma$ error of $\sim 2\%$ (see
Table~\ref{tbl_pol}). We should stress that for future GRBs observed
in better conditions the accuracy of individual measurements can be
significantly better than 1\%. The resulting polarization measurements
for GRB\,020405 are presented and discussed in Sect.~\ref{sect_disc}.

%

\section{Discussion}
\label{sect_disc}

\begin{figure}
\plotone{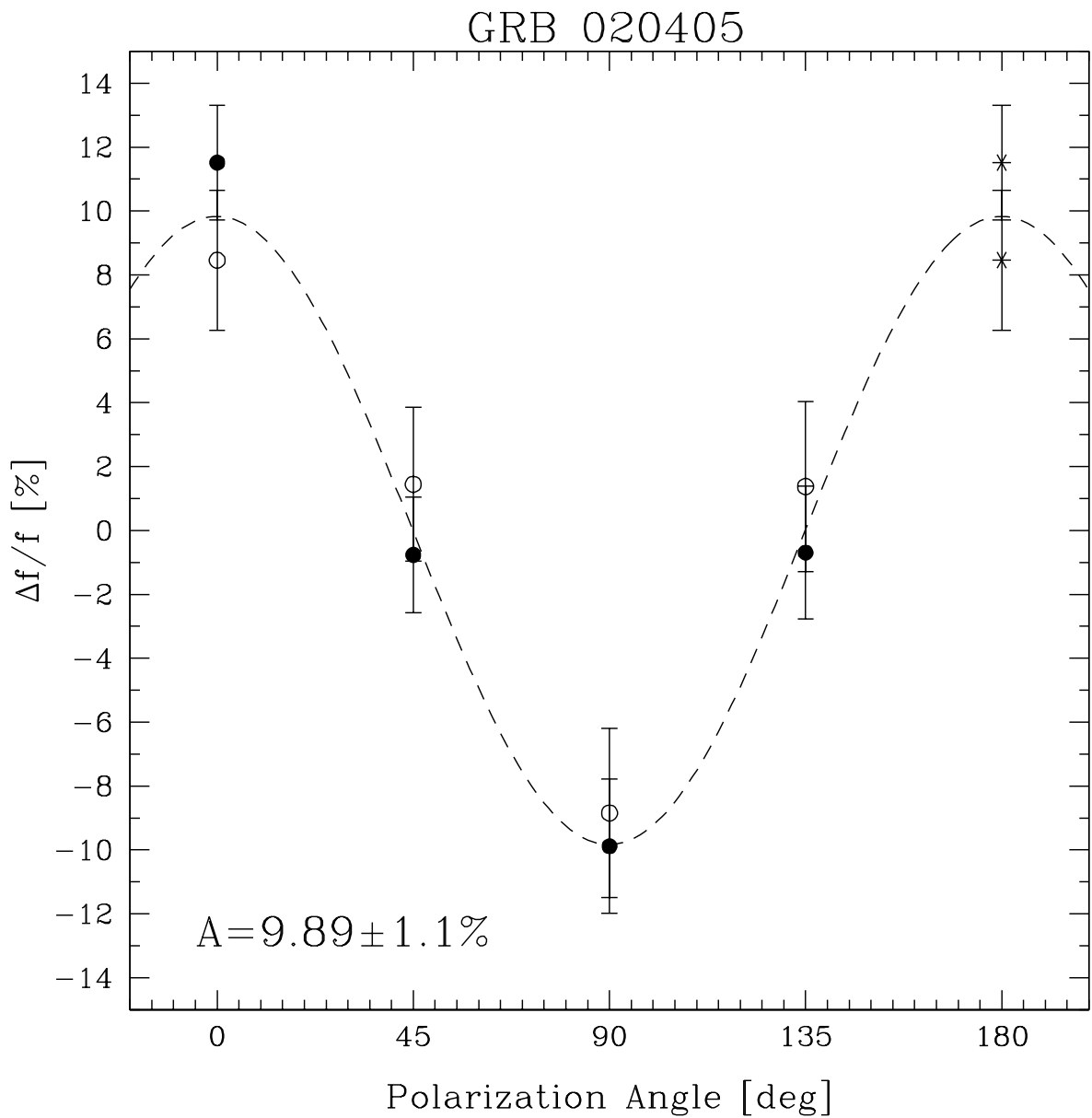}
\caption{Residual flux differences as measured through the
polarization filters. For each polarization two independent values of
flux difference were measured.  Filled circles are for the first
polarization sequence, open circles are for the second sequence. The
stars at $\theta=180\degr$ are a repeat of the points at
$\theta=0\degr$. There is a clear polarization signature present: a
sinusoid fit yields $A=9.9\pm 1.3$\% for the polarization amplitude
and $\theta=-0.1\pm 3.8 \degr$ for the polarization angle.
\label{fig_pol}}
\end{figure}

The procedure described in Section~\ref{sect_pol} resulted in
measuring the flux difference between the MiniCam observations of the
OT taken in May 2002 and those taken in April. This allows us to
obtain the total flux of the OT and also the flux differences as
measured through the polarization filters.  These flux differences,
after subtracting a slight linear trend consistent with the decay of
the OT, are shown in Fig.~\ref{fig_pol}. As two polarization sequences
were obtained, for each polarization filter two independent values of
flux difference were measured. There is a clear polarization signature
present, with the two sequences giving very similar values.  
Including the calibration uncertainty, a
sinusoid fit yields $A=9.89\pm 1.3$\% for the polarization amplitude
and $\theta=-0.1\pm 3.8 \degr$ for the polarization angle, with
resulting $\chi^2=2.3$ for six remaining degrees of freedom (compared
to $\chi^2=89.0$ if zero polarization is assumed).  Assuming a $1/A$
prior (polarization cannot be negative) yields a very similar
amplitude of $A=9.78\pm 1.3$\%. Fits to each individual sequence
yielded $A=10.7\pm 1.4$\% and $\theta = -0.4\pm 4.2 \degr$ for
the first sequence, $A=8.5\pm 1.7$\%, $\theta = 0.6\pm 6.5 \degr$
for the second sequence.

An immediate question is: could this measurement be caused by an
instrumental effect? This can be checked by constructing identical
flux difference curves for the stars in the field, which can be done
easily when using the image subtraction method.  Most of the stars in
the field have flux differences consistent with zero or relatively
small ($<3$\,\%) polarization, and no other star shows the same
polarization behavior as the OT.  In few of the cases when formally
larger ($>3$\,\%) polarization is obtained from the fit, the
polarization fits are not statistically acceptable, unlike for the
GRB. We consider the detection of large ($\sim 10$\%) polarization in
the OT of the GRB\,020405 $1.31$ days after the burst to be very
secure\footnote{To allow the astronomical community an independent
verification of this measurement, we have placed all relevant MMT
MiniCam polarization data on {\tt anonymous ftp} at {\tt
ftp://cfa-ftp.harvard.edu/pub/dbersier/GRB020405}.}.

Another question to answer is what is the chance that we would measure
this level of polarization from random photometric errors alone? We
created random flux measurements ($F_0,\ F_{45},\ F_{90},\ F_{135}$) drawn
from a a Gaussian distribution with average 1.0 and one-sigma width
0.02 (our error in magnitude). For each set of measurement we
calculate the polarization with
$$
P = 2\times \left[ \frac{(F_0 - F_{90})^2 + 
(F_{45} - F_{135})^2}{(F_0+F_{45}+F_{90}+F_{135})^2} \right]^{1/2}
$$

This has been repeated ten million times. The first result is that for
a photometric error of 0.03 mag, the most probable polarization is
$2.0$\%. The probability of having a polarization larger than 8\% from
photometric fluctuations alone is 1 in $10^7$ (for $\sigma = 0.02$
mag); it is 1 in 1000 for $\sigma = 0.03$ mag. However we have two
consecutive sequences of measurements and they are both consistent
with $>8$\% polarization. Assuming that the errors in each sequence
were 0.03 mag, the combined probability is $10^{-6}$. This does not
account for the fact that the angle is the same for each of our
observed sequences.
The inescapable conclusion is that our polarization measurement is not
due to chance photometric errors.

We calibrate the polarization to the surrounding star field so any
Galactic effect would be due to material more distant than the
surrounding stars.  Polarization could arise from the Galactic
interstellar medium; however, the foreground reddening is so small
[$E(B-V) = 0.05$] that it seems impossible that such a small amount of
dust would produce such a large polarization. Dust in the immediate
surroundings of the GRB could also induce some polarization. However,
after correcting for the small Galactic reddening, the broad-band
$BVRI$ spectrum is already very well described by a power law
(Fig.~\ref{fig_spectrum}), which is expected for a GRB. For
interstellar polarization one has $P \leq 9 E(B-V)$ \citep{smf75},
which means that a color excess $E(B-V) \sim 1.0$ would be needed to
produce the observed amount of polarization. This would produce a
intrinsic broad-band spectrum
with a heretical slope of $+1.86\pm 0.34$, and a 
very poor fit in any
case. We thus consider that dust, whether in the host galaxy or in our
Galaxy, is not the reason for the observed level of polarization.

\section{Interpretation}

The interpretation of the high polarization fraction and its apparent
rapid variation is problematic in the context of existing models.  The
polarization from the afterglow from a beamed gamma-ray burst peaks at
twenty percent for a line of sight at the very edge of the jet if the
magnetic field is restricted to the plane of the shock, so the
combination of an auspicious geometry, highly anisotropic magnetic
field, and good timing can account for the observed polarization
(Ghisellini \& Lazzati 1999).  However, the maximum polarized fraction
in a beamed gamma-ray burst is found when $\Gamma \theta \sim 1$ so
the emission will also exhibit a jet break.  The break is quite
gradual for the extreme viewing angle required by the high observed
polarization.  The spectral index and the decline rate of the flux,
when the polarization was observed, reflect emission from an isotropic
afterglow (Sari et al. 1999).  This is consistent with the
expectations of the \citet{gl99} model, but subsequent photometric
observations would reflect the emission from a jet.  We do not see any
evidence for a break in the light curve up to four days after the
burst.

Alternatively, the high observed polarization may reflect the
alignment of the magnetic field over causally connected regions in the
observed portion of the afterglow (Gruzinov and Waxman 1999).  In an
isotropic afterglow, the polarized fraction can reach ten percent if
the coherence length of the magnetic field grows at the speed of
light.  In this case, the high net polarization need not be followed
by a break in the light curve.

If the high observed polarization results from either the observation
of a beamed afterglow with $\Gamma \theta \sim 1$ or an isotropic
afterglow with a coherent magnetic field in the plane of the shock,
the timescale for the polarization fraction or angle to vary is the
time since the burst.  

Perhaps the properties of the relativistic shock in GRB afterglows is
somewhat more complicated than is commonly accepted.  The highly
variable blazars, BL Lac, OJ~287 (Angel \& Stockman 1980) and
S5~0716+714 (Impey et al. 2000), both exhibit polarized fractions up
to 20\% in the optical and rapid intra-day variability in both the
angle and extent of polarization.  The optical emission is thought to
come from a relativistic shock (with $\Gamma \sim 100$) at a distance
of $10^{12}$ to $10^{13}$~cm from the central engine.  However, in
models of BL Lacs, although changes in polarization are not generally
accompanied by changes in the flux, the fluxes vary on the same typical
timescale (Impey et al. 2000).
In the case of GRBs some short term flux variability has been observed.
For instance \citet{holland02} did see short term variability in GRB
011211 ($\sim 0.3$ mag) $\sim 0.5\;$days after the burst.  In the
present case however, any time variability would have had to be
periodic with a period very close to our observing cadence ($\sim 31$
minutes for the whole sequence). We consider it unlikely that we would
have taken our polarization measurements right when a sinusoidal flux
variation was happening.

Another scenario that can explain a temporarily high polarization is
microlensing.  \citet{lp98} predicted that polarization of a GRB may
triple during a microlensing event (increasing from 4\% to 12\%) for a
particular model, if the polarized emission emerges of a number of
coherent patches \citep{gw99}.  The polarization can also vary on a
timescale of hours, depending on the impact parameter.  \citet{lp98}
considered the polarized signature of microlensing if the GRB emission
is restricted to an annulus of ten percent thickness.  In this case,
the total emission from the burst may deviate from a power-law trend
by up to 40\%.  As the width of the ring increases, both the peak
magnification and the net polarization decrease \citep{in01}.  One
should note that polarization measurements obtained one and two nights
after our data \citep{covino02c} show that the polarization was
significantly lower, $P=1.93 \pm 0.33$\%, $\theta=154 \pm 5\degr$ on
Apr 7.21, and $P=1.23 \pm 0.43$\%, $\theta=168 \pm 9\degr$ on Apr
8.26.  To distinguish between various models will require a
polarization light curve. We can only urge other groups to publish
data (photometry and polarization) related to this burst in order to
obtain a clearer picture of this event.


\acknowledgments

We thank D. Lazzati and E. Palazzi for discussions.  DB acknowledges
support from NSF grant AST-9979812.  JK was supported by Polish KBN
grant 5P03D004-21.  JSH was supported by the Chandra Postdoctoral
Fellowship Award \#PF0-10015 issued by the Chandra X-ray Observatory
Center, which is operated by the Smithsonian Astrophysical Observatory
for and on behalf of NASA under contract NAS8-39073.

\clearpage

\begin{deluxetable}{ l l c}
\tabletypesize{\scriptsize}
\tablecaption{Observational material \label{tbl_obs}}
\tablewidth{0pt}
\tablehead{ \colhead{UT Date, UT range} & \colhead{Telescope} &
\colhead{Filters} }
\startdata
Apr 06, 06:35 -- 09:54 & VATT & $BVRI$ \\
Apr 06, 07:39 -- 09:23 & MMT$+$MiniCam    & $V$, pol. \\
Apr 08, 07:04 -- 08:56 & VATT & $VR$ \\
Apr 09, 07:41 -- 08:53 & VATT & $BVRI$ \\
Apr 09, 08:13 -- 08:36 & MMT$+$MiniCam    & $BVR$ \\
Apr 14, 07:19 -- 08:14 & Magellan$+$Magic & $V$ \\
Apr 18, 00:19 -- 00:53 & Magellan$+$LDSS2 & $R$ \\
May 2, 02:55 -- 03:17 & du Pont 100''$+$TEK5 & $BVRI$ \\
May 3, 01:48 -- 2:40 & du Pont 100''$+$TEK5 & $R$ \\
May 15 -- 16	& MMT	& $V$ \\
\enddata
\end{deluxetable}

\begin{deluxetable}{ c c r c}
\tabletypesize{\scriptsize}
\tablecaption{Polarization measurements on April 6 2002\label{tbl_pol}}
\tablewidth{0pt}
\tablehead{ \colhead{UT} & \colhead{Filter angle} & 
\colhead{$\Delta$Flux (\%)} & \colhead{$\sigma_{\Delta f}$ (\%)} }
\startdata
07:46:25 &    0	& $11.5$ & 1.8 \\
07:53:01 &   45	& $-0.8$ & 1.8 \\
07:59:07 &   90	& $-9.9$ & 2.1 \\
08:03:14 &  135	& $-0.7$ & 2.1 \\
08:18:45 &    0	& $ 8.5$ & 2.2 \\
08:25:09 &   45	& $ 1.5$ & 2.4 \\
08:31:22 &   90	& $-8.8$ & 2.6 \\
08:38:45 &  135	& $ 1.4$ & 2.7 \\
\enddata
\end{deluxetable}

\clearpage

\end{document}